# The Dynamics of Students' Behaviors and Reasoning during Collaborative Physics Tutorial Sessions

Luke D. Conlin[*], Ayush Gupta[*], Rachel E. Scherr[*], and David Hammer[†]

[*]Department of Physics, University of Maryland, College Park, MD 20742
[†]Department of Curriculum & Instruction and Department of Physics,
University of Maryland, College Park, MD 20742

**Abstract.** We investigate the dynamics of student behaviors (posture, gesture, vocal register, visual focus) and the substance of their reasoning during collaborative work on inquiry-based physics tutorials. Scherr has characterized student activity during tutorials as observable clusters of behaviors separated by sharp transitions, and has argued that these behavioral modes reflect students' epistemological framing of what they are doing, i.e., their sense of what is taking place with respect to knowledge. We analyze students' verbal reasoning during several tutorial sessions using the framework of Russ, and find a strong correlation between certain behavioral modes and the scientific quality of students' explanations. We suggest that this is due to a dynamic coupling of how students behave, how they frame an activity, and how they reason during that activity. This analysis supports the earlier claims of a dynamic between behavior and epistemology. We discuss implications for research and instruction.

## INTRODUCTION

In an analysis of students collaborating on inquiry-based physics tutorials, Scherr [1] indicated preliminary evidence of a connection between students' behaviors and the level of their scientific reasoning. Students' behaviors and reasoning during an activity are influenced by the set of expectations, i.e. 'frame,' they associate with the activity. The tutorials are meant to provide students with opportunities to practice their inquiry and reasoning skills. But is this how the students frame these activities? In the present paper we report more episodes of behavior and reasoning during tutorials, provide strong evidence for the connections between behavior, framing and reasoning, and offer reasons for such connections.

## MECHANISTIC REASONING

One of the goals of inquiry-based science teaching is to give students opportunities to learn via authentic scientific practices. Constructing scientific explanations of phenomena is a central part of authentic science, but what constitutes a scientific explanation? Based on philosophy of science literature we may adopt a tentative definition of a scientific explanation: an argument that lays out the causal mechanisms by which the phenomenon occurs, given the laws and initial conditions [2]. If constructing mechanistic explanations is an essential feature of doing science, then inquiry-based science instruction should give students opportunities to develop mechanistic reasoning skills. Indeed, inquiry in science may be described as "the pursuit of coherent mechanistic accounts of phenomena" [3].

In the present work, we use the coding scheme developed by Russ [4] to recognize and analyze mechanistic reasoning in students' verbal responses during physics tutorials. Mechanistic reasoning about a phenomenon involves several elements that Russ roughly organized into a hierarchy of increasing quality of evidence: describing target phenomenon, identifying set up conditions, identifying entities, identifying actions, identifying properties of entities, identifying the organization of entities, and "chaining." Chaining is the most complete of these elements, and involves linking several of the elements together, either to make a prediction or to reason about how things must have been in the past. A wide range of conversations produce lower level codes; mechanistic reasoning is best evidenced by chaining.

An example of chaining occurred when one group was discussing why rubbing two hands together does not produce sparks. The group had identified moisture as an entity relevant to explaining this, but could not explain further what moisture had to do with it. John [5] finally chained together an explanation based on the entity (moisture) and its properties (it is a conductor). He argued, "It's a conductor so like it's not going to let charge build on your hands because moisture's a conductor so it's like going to dissipate off into the atmosphere..."

Russ's framework allows a systematic means of assessing the substance of students' reasoning with respect to causal mechanisms, whether or not those mechanisms correspond to the "correct" account. In this paper, we examine the mechanistic reasoning of physics students during the introductory physics tutorials, and note how it is influenced by the students' framing of the tutorial activity.

## FRAMING AND BEHAVIOR

A 'frame' is the set of expectations each participant brings into a situation that corresponds with their sense of "what is going on here" [6]. For example, students in tutorial may frame the activity as a 'complete the worksheet' task, or as a 'make sense of physics' task. What counts as an appropriate explanation depends on the experience and expectations of those providing and evaluating the explanations [7]. Therefore, we should expect that student explanations of phenomena should be influenced by how they frame the activity.

Although we cannot directly observe a set of expectations, there are linguistic and behavioral cues that can inform us how a student is framing an activity. Scherr used vocal register, linguistic cues, body language, and gesture to identify four locally stable clusters of behaviors, which she calls "behavior modes," among groups of physics students working on tutorials. She initially coded the student groups for changes in behavior, using theory-neutral colors, and then used these modes as indicators of how students frame the activity. For example, the "blue" behavioral mode, in which students have their eyes on their worksheets and speak in soft voices, indicates students framing the activity as "completing the worksheet." The modes and frames are summarized in Table 1. Each behavior mode is easily observable, stable for time periods ranging from seconds to minutes, delineated by sharp transitions, and largely consistent for the entire group of students working together. For the current investigation, we use this methodology to characterize several locally coherent behavior modes, and examine the nature of explicit reasoning during these modes.

**TABLE 1.** Behavior Mode Coding Scheme

| Color Code | Behaviors | Frame |
|---|---|---|
| Green | Sitting up, eye contact with peers, subdued gestures, lower vocal register | Discussion |
| Blue | Hunched over, eyes on worksheet, low vocal register, writing | Worksheet |
| Yellow | Fidgeting, laughing, looking away, touching face/hair | Socializing |
| Red | Sitting up, eyes on TA, subdued gestures, lower vocal register | Receptive to TA |

## METHODOLOGY

This study involved 24 college students enrolled in an introductory algebra-based physics course at the University of Maryland. We transcribed 20-minute video clips of 4-student groups working during 6 tutorial sessions. The tutorials consist of 45-minute guided-inquiry sessions delivered via worksheet and attended by teaching assistants. The videos were first coded for behavior without a transcript. The behavior modes are coded somewhat holistically by considering the presence of behaviors in Table 1. Two independent coders agreed on 90% of behavior codes pre-discussion, to a precision of 5 seconds. Then the sessions were transcribed and coded for (a) instances of some level of mechanistic reasoning, and (b) instances of chaining. Two independent coders agreed on 87% of mechanistic reasoning and chaining codes before discussion. The codes for behavior, mechanistic reasoning, and chaining were then matched for each 5 seconds of the tutorial session. In the next two sections, we describe key findings from this data and provide a description of the dynamic between behavior, framing, and reasoning.

## DATA AND DISCUSSION

The percentages of mechanistic reasoning codes and of chaining specifically within each behavior mode are reported in Table 3. The pure blue, green, red, and yellow behavior modes account for nearly all of the time spent in the tutorials (86%), while the remaining time was spent in 'mixed modes.' Since the behavior modes practically span the time spent during tutorial, we will refer to the frame rather than the color code scheme for the remainder of the paper.

**TABLE 2.** Mechanistic Reasoning within Frames (summary for all 6 video clips)

| Frame | Color Code | % Time | % Mechanistic Reasoning | % Chaining |
|---|---|---|---|---|
| Animated Discussion | Green | 25% | 53% | 81% |
| Completing the Worksheet | Blue | 32% | 18% | 6% |
| Receptive to TA | Red | 24% | 15% | 4% |

Notably, the majority of mechanistic reasoning occurred during the "animated discussion" frame (53%). A significant amount of mechanistic reasoning also occurred during the "completing the worksheet" frame (18%), and the "receptive to TA" frame (15%).

The percentage of coding in any behavior mode is of course partly a result of how much students speak; students say more in the discussion frame, so there should be a higher percentage of codes by any criteria. As such, we found that the number of mechanistic reasoning codes was proportional to the number of statements made within each mode. This was not true of chaining, which leads to the finding of importance here. The number of codes for chaining in particular is disproportionately high: nearly all of the chaining occurred during the discussion frame (81%), which shows that the nature of student reasoning is different during this frame.

There are several plausible reasons for the high percentage of chaining happening during the discussion frame. Similar to what was observed by Tannen and Wallet [8], one reason is that the discussion frame was often precipitated by conflicting lines of reasoning amongst group members, who would then need to chain arguments together in order to convince each other of their explanation. For example, one group was in the worksheet frame during a tutorial involving the application of Newton's Third Law to a collision between a massive truck and a less massive car. Eric held that the forces between them are equal. Maya disagreed, saying "I don't think there's any way that you can explain to me how a massive truck is going to have the same force…" after which the group abruptly shifts into the discussion frame as they try to convince each other of their respective points of view.

While students' framing of the tutorial may influence their reasoning, it also seems the opposite may occur; the nature of explanations may shift the framing. Students in the worksheet frame tend to make comments coded at low-levels of mechanistic reasoning as they negotiate the target phenomenon to be explained and the entities involved. When a 'critical mass' of lower mechanistic reasoning codes occur, the group will transition into the discussion frame as they collectively chain together how these entities act to bring about the phenomenon. Once a sufficiently chained explanation is agreed upon, the group will shift back to the worksheet frame in order to move on with the tutorial.

The effect of reasoning on the evolution of the group's frame can be seen in graphs such as Figure 1, where a single video clip is coded from start to finish. The arrows indicate chaining, which typically occurs near transitions between worksheet and discussion frames. For this particular case, six of the nine instances of chaining occurred on the brink of such a transition.

## COUPLED DYNAMICS OF BEHAVIOR AND REASONING

Since the tutorial involves both group discussion and a worksheet, there seems to be an ever-present tension between these two ways of framing the activity. Out of the individual framing of each student emerges an overarching group dynamic that evolves over the course of the tutorial. This is evident in that all four students occupy the same behavior modes for 86% of the time.

Based on the behaviors, we have seen that the overall framing of the activity varies between tutorials and between groups. For example, Table 3 compares the framing for two groups working on different tutorials. It seems that Group 1 has more of a tendency to frame the tutorial as a discussion, while Group 2 tends to frame it more as a worksheet-completion activity. More research needs to be done to explore what influences groups to frame the tutorial in a certain way.

**TABLE 3.** Time Spent in Frames. This compares two groups working on different tutorials.

| Frame | Group 1 | Group 2 |
| --- | --- | --- |
| Discussion | 47% | 25% |
| Worksheet | 20% | 45% |
| Socializing | 4% | 14% |
| Receptive to TA | 17% | 13% |
| Other | 12% | 3% |

Each group transitions in their behavior and framing together, and these transitions are generally quite abrupt. There are many ways that the group's framing may shift. Sometimes the frames switch due

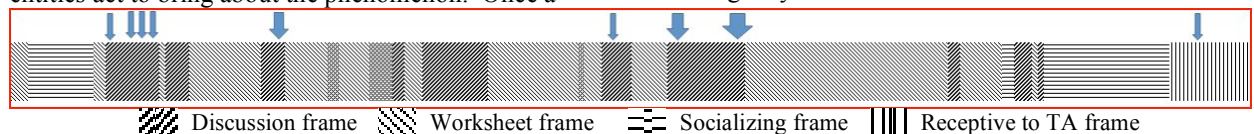

Discussion frame   Worksheet frame   Socializing frame   Receptive to TA frame

**FIGURE 1.** Tutorial Frame & Reasoning Evolution. This represents a group during a 20-minute clip from a single tutorial on Work & Energy, with time running left to right.

to an explicit frame negotiation from one or more of the members. For example, a group working on a Newton's third law tutorial had gotten off task. They were in the socializing frame, joking and talking about things that were off-topic. After a few minutes, Eric made a subtle appeal to bring the group's focus back to the worksheet, saying, "Alright, alright, alright…" while hunching over, putting his pencil near his worksheet and starting to read. The group soon transitioned into the worksheet frame.

Another reason for each group's shifts in behaviors and framing is due to contrasting student epistemologies. During their discussion about Newton's third law during a car-truck collision, Macy drew on her personal experience: "But I'm trying to think, like, okay for example I was, when I was in a car accident..." which sparks the students into a brief period of discussion. In the end Macy is not satisfied with the group's explanation that the forces are equal: "I don't know it just doesn't seem right to me." The discussion mode ends abruptly when Eric proclaims his differing epistemological stance, in which he appeals to authority: "I don't try to argue with the laws of physics. I just trust that they work."

## IMPLICATIONS AND FURTHER RESEARCH

The results here support Scherr's analysis of behavioral modes as indicative of epistemological framing. In addition, we find that the substance of student reasoning shows different patterns during different behavioral modes. Our first interest here is in developing better models of student reasoning in tutorials, but we note one instructional implication: Since behavior modes are so distinct and stable, an instructor can recognize them in real-time and act on them accordingly. For example the discussion frame is more associated with mechanistic reasoning than others, so an instructor may treat it as one sign of a well-functioning group.

The causes of frame transitions need to be examined in more detail. Specifically, the connection between chaining and frame transitions as well as the effects of the teaching assistants on the behavior modes and student reasoning should be further studied. It may also be helpful to examine students' written work in tutorial, to more accurately assess the nature of reasoning during the worksheet frame.

## SUMMARY

We have shown that there is a strong connection between how students behave during a tutorial activity and the substance of their verbal reasoning. Specifically, our evidence suggests that the most advanced mechanistic reasoning occurs while students are engaged in animated discussions. We have given several reasons for frame and behavior mode changes, including contrasting lines of reasoning, contrasting epistemologies, and the dynamic of reasoning throughout the tutorial.

## ACKNOWLEDGMENTS

This work supported in part by NSF grant REC 0440113. Further thanks go to Joe Redish, Rosemary Russ, Tom Bing, Renee-Michelle Goertzen, Brian Frank, Matty Lau, Andy Elby, and the rest of the Physics Education Research Group at the University of Maryland for their helpful comments and discussions.